# Quantum oscillations in a hexagonal boron nitride-supported single crystalline InSb nanosheet[*]

Li Zhang(张力)[1], Dong Pan(潘东)[2], Yuanjie Chen(陈元杰)[1], Jianhua Zhao(赵建华)[2], and Hongqi Xu(徐洪起)[1,3,†]

[1]*Beijing Key Laboratory of Quantum Devices, Key Laboratory for the Physics and Chemistry of Nanodevices and School of Electronics, Peking University, Beijing 100871, China*
[2]*State Key Laboratory of Superlattices and Microstructures, Institute of Semiconductors, Chinese Academy of Sciences, P.O. Box 912, Beijing 100083, China*
[3]*Beijing Academy of Quantum Information Sciences, Beijing 100193, China*

（Dated: May 19, 2022）

A gated Hall-bar device is made from an epitaxially grown, free-standing InSb nanosheet on a hexagonal boron nitride (hBN) dielectric/graphite gate structure and the electron transport properties in the InSb nanosheet are studied by gate-transfer characteristic and magnetotransport measurements at low temperatures. The measurements show that the carriers in the InSb nanosheet are of electrons and the carrier density in the nanosheet can be highly efficiently tuned by the graphite gate. The mobility of the electrons in the InSb nanosheet is extracted from low-field magneotransport measurements and a value of the mobility exceeding ~18000 $cm^2V^{-1}s^{-1}$ is found. High-field magentotransport measurements show well-defined Shubnikov-de Haas (SdH) oscillations in the longitudinal resistance of the InSb nanosheet. Temperature-dependent measurements of the SdH oscillations are carried out and key transport parameters, including the electron effective mass $m^*$~0.028 $m_0$ and the quantum lifetime $\tau$~0.046 ps, in the InSb nanosheet are extracted. It is for the first time that such experimental measurements have been reported for a free-standing InSb nanosheet and the results obtained indicate that InSb nanosheet/hBN/graphite gate structures can be used to develop advanced quantum devices for novel physics studies and for quantum technology applications.

**Keywords:** InSb nanosheet, SdH oscillations, electron effective mass, quantum lifetime
**PACS:** 85.35.Be, 73.63.-b, 73.50.Jt

---

[*]Project supported by National Key Research and Development Program of China (Grant Nos. 2017YFA0303304 and 2016YFA0300601), the National Natural Science Foundation of China (Grant Nos. 92165208, 92065106, 61974138, 11874071, 91221202, and 91421303), and the Beijing Academy of Quantum Information Sciences (Grant No. Y18G22). D. P. also acknowledges support from the Youth Innovation Promotion Association, Chinese Academy of Sciences (Grant Nos. 2017156 and Y2021043).
[†]Corresponding author. Email: hqxu@pku.edu.cn

# 1. Introduction

In recent years, narrow bandgap, III-V binary compound semiconductor InSb has attracted great attention due to its light electron effective mass, high electron mobility, strong spin-orbit interaction, and large Landé $g$-factor[1-4] and its appealing applications in high-speed electronics,[5] spintronics,[6] quantum electronics,[7] and topological quantum computation.[8] Especially, in the emerging field of topological quantum computation, gate-controlled InSb nanowire-superconductor hybrid devices were the first solid state systems in which signatures of Majorana fermions were detected.[9-11] However, to build topological devices, in which braiding of Majorana fermions can be conveniently performed, a move from single straight nanowire structures to branched nanowire,[12] multiple nanowire,[13,14] and/or two-dimensional (2D) planar quantum structures,[15,16] could be inevitably required. Recently, high quality InSb quantum wells (QWs)[17-20] and free-standing nanosheets[21-25] have been realized with epitaxial growth techniques and have been characterized by low-temperature quantum transport measurements. In these characterization measurements, well-defined Shubnikov-de Haas (SdH) oscillations, which arise due to the Landau level formation, have been observed in the InSb QW structures and have been employed as an effective means to quantify the transport properties of these semiconductor planar quantum structures.[17,18,20] In strong contrast, observations of SdH oscillations with a quality sufficient for extraction of quantum transport properties in the emerging epitaxially-grown free-standing InSb nanosheets are however still lacking, even though these nanosheets would have an advantage of making direct metal contacts on their surfaces and various types of functional devices, such as superconducting Josephson junctions[26-28] and quantum dots[29,30], have been realized with these InSb nanosheets. The difficulty in observing conspicuous intrinsic quantum oscillations is most likely due to the fact that the InSb nanosheet devices have so far been made on $SiO_2$/Si substrates[25] and thus the electrons in the nanosheets have suffered strong scattering from charged impurities and defects in the $SiO_2$ dielectrics and at the interfaces between the InSb nanosheets and the $SiO_2$ dielectrics[31-35].

With advances in experimental techniques for aligning/stacking 2D materials,[32,36-39] hexagonal boron nitride (hBN) flakes have been widely used as supporting and/or encapsulation layers in device fabrication. It has been experimentally demonstrated that the devices fabricated in this way have significantly improved the transport properties of the 2D materials when compared to their counterpart devices fabricated directly on SiO$_2$/Si substrates.[40-46] This motivates us to consider stacking epitaxially grown, free-standing InSb nanosheets on exfoliated hBN layers, instead of directly on SiO$_2$/Si substrates, for the fabrication of high-quality Hall-bar devices and to see whether well-defined quantum oscillations could be observed in these free-standing InSb nanosheet devices.

In this paper, we report on the realization of a good quality InSb Hall-bar device from an InSb nanosheet/hBN/graphite trilayer heterostructure. The InSb nanosheet employed as a conduction channel in the device is a free-standing, micrometer-sized layer and is obtained via molecular beam epitaxy (MBE). The hBN and graphite layers are obtained by exfoliation and are used as the dielectric and the gate electrode to the InSb nanosheet channel, respectively. The fabricated device is characterized by gate-transfer characteristic and magnetotransport measurements at low temperatures. The measurements demonstrate that the device exhibits an excellent gate control of the electron density in the InSb nanosheet and the Hall mobility in the nanosheet is exceeding $1.8 \times 10^4$ cm$^2$ V$^{-1}$ s$^{-1}$. Such a high mobility enables us to observe well-defined SdH oscillations in the measured longitudinal resistance and to carry out a detailed analysis of the quantum oscillations in a free-standing InSb nanosheet for the first time. We have performed the measurements of the SdH oscillations at different temperatures. The key electron transport properties, such as the electron effective mass ($m^*$~0.028 $m_0$) and the quantum lifetime ($\tau$~0.046 ps), in the InSb nanosheet are extracted from the temperature-dependent SdH oscillation amplitudes.

## 2. Experimental methods

### 2.1. Materials and fabrication of an InSb nanosheet/hBN/graphite trilayer

The InSb nanosheet Hall-bar device was made from a planar InSb nanosheet/hBN/graphite trilayer heterostructure placed on a SiO$_2$/Si substrate. The micrometer sized InSb nanosheet in the trilayer was grown on an InAs nanowire by MBE on a Si(111) substrate in a free-standing form and was a zincblende crystal.[21] Figure 1(a) shows a scanning electron microscope (SEM) image of an as-grown InSb nanosheet on the same growth substrate as the InSb nanosheet used in this work. Figure 1(b) shows a high-resolution transmission electron microscope (HRTEM) image of an InSb nanosheet from this growth substrate and the inset in Fig. 1(b) shows its fast Fourier transform (FFT) pattern. It is seen that the InSb nanosheet was a pure zincblende crystal with a flat (011) top surface and was free from stacking faults and twinning defects. The distance between two adjacent atomic layers in the nanosheet was determined from the FFT pattern to be ~0.217 nm. Atomic force microscopy (AFM) measurements showed that the thicknesses of InSb nanosheets grown by MBE in a Si(111) are in a range from 10 to 100 nm.[21]

For device fabrication, as-grown InSb nanosheets were transferred from the growth substrate to a $SiO_2$/Si substrate with predefined metal markers on top. The flakes of hBN and graphite were obtained by mechanical exfoliation and were placed on other two marker-predefined $SiO_2$/Si substrates. For making an InSb nanosheet/hBN/graphite trilayer heterostructure, an InSb nanosheet and an hBN flake were first successively picked up to form an InSb nanosheet/hBN bilayer using a home-made optical alignment and mechanical transfer setup mounted with a polypropylene carbonate (PPC) film coated stamp. This bilayer was then aligned and released onto a graphite flake.[36-39] It is important to note that since the InSb nanosheet was of about a micrometer size and was hardly seen through the optical microscope employed in the alignment/transfer setup, a pattern that duplicated the metal markers on the InSb nanosheet substrate was imprinted on the PPC film during the nanosheet pick up and was used to align and stack the InSb nanosheet on the hBN flake. Figure 2(a) shows an optical image of the InSb nanosheet/hBN/graphite trilayer heterostructure obtained and used in this work, in which the hBN and graphite flakes are marked by red and purple dotted lines, respectively. The area where the InSb nanosheet is located is marked by a black square.

## 2.2. Hall-bar device fabrication and measurement setups

In fabrication of the Hall-bar device, the InSb nanosheet/hBN/graphite trilayer heterostructure was located against the predefined markers on the substrate by optical microscope and then smaller markers were fabricated by electron beam lithography (EBL), electron beam evaporation (EBE) of Ti/Au (5/45 nm), and lift-off [see Fig. 2(a) for fabricated small markers]. Subsequently, the InSb nanosheet was located with respect to these small markers by SEM, and source and drain contacts (contacts 1 and 6), four voltage probe contacts (contacts 2-5), and a contact to the bottom graphite gate were fabricated via a combined process of EBL, EBE of Ti/Au (10/120 nm), and lift-off. We note that before contact metal deposition, the exposed areas on the InSb nanosheet were chemically etched in a de-ionized water-diluted $(NH_4)_2S_x$ solution at a temperature of 40 °C for two minutes to remove the surface oxide. Figure 2(b) shows a false-colored SEM image of the fabricated Hall-bar device, in which the InSb nanosheet has a thickness of $d$~20 nm and a width of $W$~2.5 μm, while the conduction channel length, defined by the separation between source and drain contacts 1 and 6, is $L$ ~ 2 μm. The separation between probe contacts 2 and 3 is $L_{23}$ ~ 1.4 μm and the separation between probe contacts 3 and 5 is $W_{35}$ ~ 2 μm.

The low-temperature transport measurements were conducted in a physical property measurement system (PPMS) cryostat equipped with a uniaxial magnet. In the gate-transfer characteristic measurements, a fixed DC voltage ($V_{ds}$) of 1 mV was applied to the source and drain contacts (contacts 1 and 6) and the channel current ($I_{ds}$) was detected as a function of the bottom gate voltage $V_{BG}$, see the inset of Fig. 2(c) for the circuit setup. The source-drain bias voltage $V_{ds}$ and the bottom gate voltage $V_{BG}$ were applied by voltage sources (GS200). The magnetotransport measurements were performed at the magnetic field $B$ applied perpendicular to the InSb nanosheet plane using a standard lock-in technique, in which a 17-Hz AC channel current $I_x$ of ~ 50 nA was supplied between source and drain contacts 1 and 6, and the longitudinal voltage drop $V_x$ between probe contacts 2 and 3 and the transverse Hall voltage drop $V_y$ between probe contacts 5 and 3 were recorded [see the circuit setup in Fig. 2(b)]. The excitation current and voltage drops were supplied by and detected with lock-in amplifiers (SR830) at a time constant of 1 s. In the following, we report the results all obtained from the measurements of the Hall-bar device as shown in Fig. 2(b).

## 3. Results and discussion

### 3.1. Gate-transfer characteristic measurements

Figure 2(c) shows the gate-transfer characteristic measurements of the Hall-bar device shown in Fig. 2(b) at a temperature of $T = 1.9$ K. The measurements show that the carriers in the InSb nanosheet are of $n$-type at positive gate voltages $V_{BG}$ and that the InSb nanosheet/hBN/graphite heterostructure device is a field-effect transistor operated on an enhancement mode—the conduction channel is in a pinch-off state at $V_{BG} = 0$ V and can be switched on by applying a sufficiently large positive $V_{BG}$. Assuming that there exists a contact resistance $R_c$ in the device and the field-effect mobility $\mu_{FE}$ is independent of gate voltage $V_{BG}$, the device conductance can then be expressed as[47]

$$G(V_{BG}) = (R_c + \frac{L^2}{\mu_{FE} C (V_{BG}-V_{th})})^{-1}, \qquad (1)$$

where $V_\text{th}$ is the pinch-off threshold voltage and $C = \frac{\varepsilon_0 \varepsilon_\text{r}}{t} LW$ is the gate-to-channel capacitance with $\varepsilon_0$ denoting the vacuum permittivity, $\varepsilon_\text{r} = 2.9$ the relative dielectric constant of hBN, and $t$ the thickness of the hBN flake. As determined by AFM measurements, the thickness of the hBN flake in the device is $t \approx 27$ nm. By fitting the measured on-state transfer characteristic curve to Eq. (1), $R_\text{c} \sim 1.01$ kΩ, $V_\text{th} \sim 0.92$ V and $\mu_\text{FE} \sim 7400$ cm$^2$ V$^{-1}$ s$^{-1}$ are extracted. The 2D carrier density in the InSb nanosheet can be estimated from $n = C_\text{gs} \times \frac{V_\text{BG} - V_\text{th}}{e}$, where $e$ is the elementary charge and $C_\text{gs} = \frac{\varepsilon_0 \varepsilon_\text{r}}{t} \sim 950$ μF m$^{-2}$ the unit-area gate-to-channel capacitance, and is found to be modulated from $n = 1.0 \times 10^{11}$ cm$^{-2}$ to $n = 1.2 \times 10^{12}$ cm$^{-2}$ in the linear region (1 V $\leq V_\text{BG} \leq$ 3 V) of the transfer characteristic curve [Fig. 2(d)]. This result gives a relatively large slope of $dn/dV_\text{BG} = 5.5 \times 10^{15}$ m$^{-2}$ V$^{-1}$, indicating that the thin-hBN/graphite bilayer can be used as an effective means to control the electron density in the InSb nanosheet. The electron mean free path in the nanosheet can be obtained from $L_\text{e} = \frac{\hbar \mu_\text{FE}}{e} \sqrt{2\pi n}$ (where $\hbar$ is the reduced Planck constant) and is found to be $L_\text{e} \sim 135$ nm at $n = 1.2 \times 10^{12}$ cm$^{-2}$. This mean free path value is much smaller than the channel length and thus the electron transport in the nanosheet is in the diffusive regime at zero magnetic field.

### 3.2. Low-field magnetotransport measurements

Figure 3 shows a series of low-field magnetotransport measurements at $T = 1.9$ K. Figure 3(a) shows the Hall resistance ($R_\text{yx} = V_\text{y}/I_\text{x}$) as a function of magnetic field $B$ at different gate voltages $V_\text{BG}$. The measured $R_\text{yx}$-$B$ curve at each $V_\text{BG}$ can be excellently fitted by a line and the Hall coefficient $R_\text{H}$ can be obtained from the slope of the fitting line. For example, at $V_\text{BG} = 2.5$ V, a Hall coefficient value of $R_\text{H} \sim 720$ Ω/T can be extracted from the Hall resistance measurements, which gives a sheet carrier density of $n_\text{sheet} \sim 8.7 \times 10^{11}$ cm$^{-2}$. The carrier densities $n_\text{sheet}$ in the InSb nanosheet at other $V_\text{BG}$ can be obtained similarly. Figure 3(c) shows the values (red dots) of sheet carrier density $n_\text{sheet}$ determined for the InSb nanosheet at different $V_\text{BG}$. It is seen that $n_\text{sheet}$ is increased linearly from $7.5 \times 10^{11}$ cm$^{-2}$ to $1.0 \times 10^{12}$ cm$^{-2}$ with increasing $V_\text{BG}$. The red dashed line in Fig. 3(c) shows a line fit of the $n_\text{sheet}$ data points. The slope of the fitting line gives a unit-area gate-to-nanosheet capacitance of ~ 956 μF m$^{-2}$, which is very close to the estimated value of ~ 950 μF m$^{-2}$ obtained above. From the determined sheet carrier densities $n_\text{sheet}$ shown in Fig. 3(c), the three-dimensional (3D) bulk electron density $n_{3D}$ at different $V_{BG}$ can be estimated out by normalizing $n_{sheet}$ with the thickness of the InSb nanosheet ($d\sim20$ nm). The bulk Fermi wave vectors $k_F = (3\pi^2 n_{3D})^{1/3}$ and Fermi wavelengths $\lambda_F$ in the InSb nanosheet can then be obtained. The results are $k_F \sim 0.226$ nm$^{-1}$ and $\lambda_F \sim 27.8$ nm at $V_{BG} = 2.35$ V, $k_F \sim 0.229$ nm$^{-1}$ and $\lambda_F \sim 27.4$ nm at $V_{BG} = 2.4$ V, $k_F \sim 0.234$ nm$^{-1}$ and $\lambda_F \sim 26.8$ nm at $V_{BG} = 2.5$ V, and $k_F \sim 0.239$ nm$^{-1}$ and $\lambda_F \sim 26.3$ nm at $V_{BG} = 2.6$ V. It is seen that all these determined Fermi wavelengths are larger than the thickness of the InSb nanosheet and thus the electron transport in the nanosheet is in the two-dimensional (2D) regime.

Figure 3(b) shows the measured longitudinal resistance ($R_{xx} = V_x/I_x$) at different $V_{BG}$ within the same small magnetic field region as in Fig. 3(a). The measurements show clear universal conductance fluctuations due to quantum interference. The electron Hall mobility $\mu_{Hall}$ can be estimated from the measured longitudinal resistance values of $R_{xx}(0)$ at zero magnetic field via

$$\mu_{Hall} = \frac{L_{23}}{W} \frac{1}{R_{xx}(0) n_{sheet} e}. \qquad (2)$$

The black squares in Fig. 3(c) are the Hall mobilities $\mu_{Hall}$ obtained using $L_{23} = 1.4$ μm and $W = 2.5$ μm in Eq. (2). It is seen that the Hall mobility is increased from 15100 cm$^2$ V$^{-1}$ s$^{-1}$ with increasing $V_{BG}$ from $V_{BG}$=2.3 V and stays at a high value of 18200 cm$^2$ V$^{-1}$ s$^{-1}$ at $V_{BG}$=2.5 V. After $V_{BG}$=2.5 V, the Hall mobility is slightly decreased with further increasing $V_{BG}$. This slight decrease in Hall mobility with the continuous increase in $V_{BG}$ is presumably due to the fact that at these high gate voltages, significantly more electrons in the nanosheet are attracted to locate against the bottom nanosheet surface and thus suffer stronger scattering from defects, impurities and charge centers at the surface and in the native nanosheet oxidized layer. Now, it is important to note that all these extracted values of the Hall mobility shown in Fig. 3(c) are much larger than the field-effect mobility $\mu_{FE}$ obtained above from the gate-transfer characteristic measurements. This significant difference arises due to the fact that the field-effect mobility is assumed to be independent of gate voltage (i.e., carrier density) and represents an average value over a large range of on-state gate voltages. Thus, for a precise characterization of the mobility in a layer structure, it is important to construct a Hall-bar device and carry out standard low-field magnetotransport measurements. We should also note that these Hall mobility values presented in Fig. 3(c) are most likely underestimated values because the Hall-bar device was made with the voltage probes covering parts of the nanosheet channel which could introduce additional scattering to the electrons in the nanosheet channel.

We note that the extracted Hall mobility in the InSb nanosheet has not reached a value achieved in a heterostructure-defined InSb QW.[17-20] Several limiting factors for the mobility in the device could be considered. But, the influence of structural defects in the InSb nanosheet lattice can be eliminated as the nanosheet was grown by MBE with a high crystalline quality as discussed above [see Fig. 1(b)]. Similarly, scattering due to charged impurities in the $SiO_2$ dielectric has most likely been screened out by the graphite gate and the charge trap-free hBN dielectric.[32,45] The primary limiting factor for the mobility in the InSb nanosheet is most likely the scattering by defects and impurities in the native nanosheet oxidized layers and at the nanosheet-oxidized layer interfaces[48-50]. Thus, a further technology development for removal of InSb nanosheet surface oxide layers and assembly of an InSb nanosheet on an hBN layer inside a glove box is required to obtain an exceptional mobility in the InSb nanosheet.

We note also that in this work, a relatively thinner InSb nanosheet was selected to make the InSb nanosheet/hBN/graphite trilayer Hall-bar device in order to extract 2D transport properties of electrons in a free-standing semiconductor nanosheet. If a thicker InSb nanosheet was used in the study, the effect of the native oxidized layers of the nanosheet would get smaller and thus the mobility would get increased. However, in a thicker InSb nanosheet, the electron transport at a large positive gate voltage could easily enter a quasi-2D regime, in which the electrons populate more than one 2D subbands, leading to a decrease in the electron mobility due to occurrence of inter-2D subband scattering. In addition, the electron occupation of more than one 2D subbands will make observation of a single set of well-defined SdH oscillations difficult, leading to a difficulty in achieving an accurate analysis of the temperature dependent measurements of SdH oscillations using the Lifshitz-Kosevich (L-K) formula[40] for extraction of electron transport parameters.

### 3.3. Gate dependent measurements of the quantum oscillations

Figure 4(a) shows the longitudinal resistance $R_{xx}$ and the Hall resistance $R_{yx}$ measured for the Hall-bar device shown in Fig. 2(b) at $V_{BG}$ = 2.5 V and $T$ = 1.9 K at magnetic fields $B$ applied up to 5.5 T. The SdH oscillations are seen to appear at magnetic fields $B$ > 2 T, which is consistent with the high electron mobility determined above in the InSb nanosheet. The black arrows in Fig. 4(a) mark the magnetic field positions at the bottoms of a few consecutive SdH oscillation valleys (to which the corresponding even values of filling factors $v$ are assigned). Figure 4(b) shows $R_{xx}$ as a function of $B$ measured for the Hall-bar device at several gate voltages $V_{BG}$ and at $T = 1.9$ K. Note that all the measured curves except for the one obtained at $V_{BG} = 2.35$ V are shifted vertically for clarity. The short black arrows in Fig. 4(b) again mark the magnetic field positions $B_v$ at the bottoms of a few consecutive SdH oscillation valleys at which the corresponding even filling factors from $v = 8$ to $v = 14$ can be assigned. The inset of Fig. 4(b) shows plots of $1/B_v$ as a function of the assigned $v$ at different $V_{BG}$ [data points in the inset of Fig. 4(b)]. The dashed lines in the inset of Fig. 4(b) are the linear fits to the measured data points. Here, it is seen that excellent linear fits to the data points are obtained at all considered $V_{BG}$. The slopes of the linear fits give the oscillation periods $\Delta(1/B)$. As expected, a smaller period is observed at a higher gate voltage or a higher carrier density. At the four considered values of $V_{BG} = 2.35$, 2.4, 2.5, and 2.6 V, the oscillation periods are determined to be $\Delta(1/B) = 0.0624$, 0.0614, 0.0568, and 0.0537 T$^{-1}$, respectively. According to the relation of $\Delta(1/B)=(2\pi)^2 e/hS_F$, where $h$ is the Planck constant and $S_F$ is the cross-sectional area of the Fermi surface in a plane perpendicular to the magnetic field, $S_F$ can be obtained as $S_F = 0.152$, 0.155, 0.167, and 0.177 nm$^{-2}$ at the four considered values of $V_{BG}$. The corresponding planar (2D) Fermi wave vectors and Fermi wavelengths can then be estimated to be $k_F \sim 0.220$ nm$^{-1}$ and $\lambda_F \sim 28.6$ nm at $V_{BG} = 2.35$ V, $k_F \sim 0.222$ nm$^{-1}$ and $\lambda_F \sim 28.3$ nm at $V_{BG} = 2.4$ V, $k_F \sim 0.230$ nm$^{-1}$ and $\lambda_F \sim 27.3$ nm at $V_{BG} = 2.5$ V, and $k_F \sim 0.237$ nm$^{-1}$ and $\lambda_F \sim 26.5$ nm at $V_{BG} = 2.6$ V. The Fermi wave vectors are slightly smaller than their corresponding values obtained above from the low-field magnetotransport measurements. This is consistent with the fact that the effect of magnetic field depletion of electrons in the nanosheet becomes noticeable at high magnetic fields.

**3.4. Temperature dependent measurements of the quantum oscillations**

In order to extract more transport parameters from the SdH oscillations of electrons in the InSb nanosheet, such as cyclotron effective mass, Fermi velocity and quantum lifetime, we analyze the evolution of the SdH oscillation amplitudes with changes in temperature and magnetic field. Figure 5(a) shows the measured longitudinal resistance $\Delta R_{xx}$, after subtracting a smooth background, as a function of $1/B$ at different temperatures and at $V_{BG} = 2.5$ V. Here, vertical dashed lines in the figure mark the magnetic field positions of a few consecutive peaks and valleys, where the SdH oscillation amplitudes decrease with increasing temperature. Figure 5(b) shows the SdH oscillation amplitudes $|\Delta R_{xx}|$ (solid symbols) extracted from Fig. 5(a) as a function of temperature at these oscillation peaks and valleys (which are identified by their magnetic field position values given on the left side of the curves). The cyclotron effective mass $m^*$ of the electrons in the InSb nanosheet can be obtained by fitting these temperature dependent amplitudes $|\Delta R_{xx}|$ to the L-K formula,[40] $A(T) \propto \lambda(T)/\sinh\lambda(T)$, where $\lambda(T) = 2\pi^2 k_B T m^*/\hbar eB$, and $k_B$ is the Boltzmann constant. The solid lines in Fig. 5(b) represent the fits to the data points. The cyclotron effective mass values extracted from the fits are shown in Fig. 5(c). It is seen that $m^*$ has a value of $m^* = 0.0247\, m_0$ extracted from the SdH oscillation amplitudes at $B = 2.47$ T, a value of $m^* = 0.0317\, m_0$ from the SdH oscillation amplitudes at $B = 2.67$ T, and values in between [see Fig. 5(c)], where $m_0$ is the bare electron mass. These extracted effective mass values give an average value of $m^* = 0.028\, m_0$ in the InSb nanosheet, which is similar to the values extracted from the measurements for InSb QWs,[17] but larger than the $m^*$ value in bulk InSb (black solid line). This discrepancy from the bulk value is as expected and is due to the quantum confinement which leads to an increase in the bandgap in the InSb nanosheet and thus an increase in the value of $m^*$.[20] From the extracted average electron effective mass value and the Fermi wave vector obtained from the low-filed magnetotransport measurements, the 2D Fermi velocity and Fermi level in the InSb nanosheet are estimated to be $v_F = \hbar k_F/m^* = 9.64 \times 10^5$ m/s and $E_F = \frac{1}{2}m^* v_F^2 = 74$ meV at $V_{BG} = 2.5$ V.

The quantum lifetime $\tau$ of the electrons in the InSb nanosheet can be obtained by analyzing the Dingle factor $R_D = e^{-D}$, where $D = \pi m^* \Gamma/eB$ with $\Gamma = 1/\tau$ denoting the scattering rate.[51] Figure 5(d) shows a plot of $\ln[|\Delta R|B\sinh\lambda(T)]$ against $1/B$ (a Dingle plot) at different temperatures (hollow symbols). The dashed lines are linear fits of the data points. Because the oscillation amplitude $A(T) \propto R_D \lambda(T)/\sinh\lambda(T)$,[52] we can extract quantum lifetime from the slope of the linear fits at different temperatures, and the results are shown in the inset of Fig. 5(d). From these results, an averaged quantum lifetime $\tau \sim 0.046$ ps is extracted for the InSb nanosheet at $V_{BG} = 2.5$ V. The Drude scattering time $\tau_D = 0.289$ ps can be calculated using the Drude model $\mu_{Hall} = e\tau_D/m^*$ at this gate voltage. The dominant scattering angles of the electrons in the InSb nanosheet can be found from the Dingle ratio $\tau_D/\tau$. While $\tau_D$ is weighted toward large-angle scattering events, $\tau$ reflects the influence of all small- and large-angle scattering mechanisms. Thus, isotropic short-range scattering can be expected if the Dingle ratio is close to 1. Otherwise, if the Dingle ratio is remarkably greater than 1, the main scattering mechanism is from long-range Coulomb interaction.[53] Since the Dingle ratio of $\tau_D/\tau \approx 6$ is found here in this work, the small-angle scattering led by long-range Coulomb interaction is expected to play a dominant role in quantum transport in our InSb nanosheet/hBN/graphite heterostructure Hall-bar device.

## 4. Conclusions

In conclusion, a gated InSb nanosheet Hall-bar device has been realized from an InSb nanosheet/hBN/graphite trilayer placed on a $SiO_2$/Si substrate and is studied by gate-transfer characteristic and magnetotransport measurements. The InSb nanosheet is grown via MBE and is a free-standing, single-crystalline zincblende crystal. The hBN flake and the graphite flake are obtained by exfoliation. The InSb nanosheet/hBN/graphite trilayer is made from the MBE-grown InSb nanosheet, exfoliated hBN and graphite flakes using a home-made alignment/transfer setup and is fabricated into an InSb nanosheet Hall-bar device where the hBN and graphite flakes are used as the dielectric and the bottom gate. The gate-transfer characteristic and low-field magnetotransport measurements at low temperatures show that the carrier density in the InSb nanosheet can be highly efficiently tuned by the graphite gate and the carriers in the InSb nanosheet are of *n*-type at positive gate voltages. The mobility of the electrons in the InSb nanosheet extracted from these measurements reaches a high value of ~18000 $cm^2V^{-1}s^{-1}$, indicating an advantage of using the graphite gate in the device. High-field magentotransport measurements show well-defined SdH oscillations in the longitudinal resistance of the InSb nanosheet. Temperature-dependent measurements of the SdH oscillations are carried out and key transport parameters, including the electron effective mass $m^*$~0.028 $m_0$ and the quantum lifetime $\tau$~0.046 ps, in the InSb nanosheet are extracted. It is for the first time that such measurements have been carried out for an epitaxially grown, free-standing InSb nanosheet and the results obtained would form a solid basis for the development of advanced quantum devices based on InSb nanosheet/hBN/graphite trilayers for novel physics studies and for quantum technology applications.

## Acknowledgments

The authors thank Wei Sun, Ning Kang and Zixuan Yang at the Institute of Physical Electronics of Peking University for assistance with the AFM characterization measurements.

# References

[1] Gilbertson A M, Kormanyos A, Buckle P D, Fearn M, Ashley T, Lambert C J, Solin S A and Cohen L F 2011 *Appl Phys Lett* **99** 242101

[2] Kallaher R L, Heremans J J, Goel N, Chung S J and Santos M B 2010 *Physical Review B* **81** 075303

[3] Nedniyom B, Nicholas R J, Emeny M T, Buckle L, Gilbertson A M, Buckle P D and Ashley T 2009 *Physical Review B* **80** 125328

[4] Chen Y, Huang S, Pan D, Xue J, Zhang L, Zhao J and Xu H Q 2021 *npj 2D Materials and Applications* **5**

[5] Ahish S, Sharma D, Vasantha M H and Kumar Y B N 2016 *IEEE Computer Society Annual Symposium on VLSI*, 105-109

[6] Yang Z, Heischmidt B, Gazibegovic S, Badawy G, Car D, Crowell P A, Bakkers E and Pribiag V S 2020 *Nano Lett* **20** 3232

[7] Nilsson H A, Caroff P, Thelander C, Larsson M, Wagner J B, Wernersson L E, Samuelson L and Xu H Q 2009 *Nano Lett* **9** 3151

[8] Nayak C, Simon S H, Stern A, Freedman M and Das Sarma S 2008 *Reviews of Modern Physics* **80** 1083

[9] Mourik V, Zuo K, Frolov S M, Plissard S R, Bakkers E P A M and Kouwenhoven L P 2012 *Science* **336** 1003

[10] Rokhinson L P, Liu X and Furdyna J K 2012 Nature Physics 8 795

[11] Deng M, Yu C, Huang G, Larsson M, Caroff P and Xu H 2012 *Nano letters* **12** 6414

[12] Gazibegovic S, Car D, Zhang H, *et al*. 2017 *Nature* **548** 434

[13] Sau J D, Clarke D J and Tewari S 2011 *Physical Review B* **84** 094505

[14] van Heck B, Akhmerov A R, Hassler F, Burrello M and Beenakker C W J 2012 *New Journal of Physics* **14** 035019

[15] Aasen D, Hell M, Mishmash R V, Higginbotham A, Danon J, Leijnse M, Jespersen T S, Folk J A, Marcus C M, Flensberg K and Alicea J 2016 *Physical Review X* **6** 031016

[16] Hell M, Leijnse M and Flensberg K 2017 *Phys Rev Lett* **118** 107701

[17] Lehner C A, Tschirky T, Ihn T, Dietsche W, Keller J, Fält S and Wegscheider W 2018 *Physical Review Materials* **2** 054601

[18] Lei Z, Lehner C A, Cheah E, Karalic M, Mittag C, Alt L, Scharnetzky J, Wegscheider W, Ihn T and Ensslin K 2019 *Appl. Phys. Lett.* **115** 012101

[19] Ke C T, Moehle C M, de Vries F K, Thomas C, Metti S, Guinn C R, Kallaher R, Lodari M, Scappucci G, Wang T T, Diaz R E, Gardner G C, Manfra M J and Goswami S 2019 *Nat Commun* **10** 3764

[20] Lei Z, Lehner C A, Rubi K, Cheah E, Karalic M, Mittag C, Alt L, Scharnetzky J, Märki P and Zeitler U 2020 *Phys. Rev. Research* **2** 033213

[21] Pan D, Fan D X, Kang N, Zhi J H, Yu X Z, Xu H Q and Zhao J H 2016 *Nano Lett.* **16** 834

[22] de la Mata M, Leturcq R, Plissard S R, Rolland C, Magen C, Arbiol J and Caroff P 2016 *Nano Lett* **16** 825

[23] Gazibegovic S, Badawy G, Buckers T L J, Leubner P, Shen J, de Vries F K, Koelling S, Kouwenhoven L P, Verheijen M A and Bakkers E 2019 *Adv Mater* **31** 1808181
[24] Verma I, Salimian S, Zannier V, Heun S, Rossi F, Ercolani D, Beltram F and Sorba L 2021 *ACS Appl. Nano Mater.* **4** 5825
[25] Kang N, Fan D, Zhi J, Pan D, Li S, Wang C, Guo J, Zhao J and Xu H 2019 *Nano Lett.* **19** 561
[26] Zhi J, Kang N, Li S, Fan D, Su F, Pan D, Zhao S, Zhao J and Xu H 2019 *physica status solidi (b)* **256** 1800538
[27] Zhi J, Kang N, Su F, Fan D, Li S, Pan D, Zhao S P, Zhao J and Xu H Q 2019 *Physical Review B* **99** 245302
[28] de Vries F K, Sol M L, Gazibegovic S, Veld R L M o h, Balk S C, Car D, Bakkers E P A M, Kouwenhoven L P and Shen J 2019 *Phys. Rev. Research* **1** 032031
[29] Xue J, Chen Y, Pan D, Wang J-Y, Zhao J, Huang S and Xu H Q 2019 *Applied Physics Letters* **114** 023108
[30] Chen Y, Huang S, Mu J, Pan D, Zhao J and Xu H Q 2021 *Chin. Phys. B* **30** 128501
[31] Hwang E H, Adam S and Sarma S D 2007 *Phys Rev Lett* **98** 186806
[32] Dean C R, Young A F, Meric I, Lee C, Wang L, Sorgenfrei S, Watanabe K, Taniguchi T, Kim P, Shepard K L and Hone J 2010 *Nat Nanotechnol* **5** 722
[33] Chen K, Wan X, Liu D, Kang Z, Xie W, Chen J, Miao Q and Xu J 2013 *Nanoscale* **5** 5784
[34] Cui X, Lee G H, Kim Y D, Arefe G, Huang P Y, Lee C H, Chenet D A, Zhang X, Wang L, Ye F, Pizzocchero F, Jessen B S, Watanabe K, Taniguchi T, Muller D A, Low T, Kim P and Hone J 2015 *Nat Nanotechnol* **10** 534
[35] Joo M K, Moon B H, Ji H, Han G H, Kim H, Lee G, Lim S C, Suh D and Lee Y H 2017 *ACS Appl Mater Interfaces* **9** 5006
[36] Wang L, Meric I, Huang P, Gao Q, Gao Y, Tran H, Taniguchi T, Watanabe K, Campos L and Muller D 2013 *Science* **342** 614
[37] Castellanos-Gomez A, Buscema M, Molenaar R, Singh V, Janssen L, van der Zant H S J and Steele G A 2014 *2D Mater.* **1** 011002
[38] Pizzocchero F, Gammelgaard L, Jessen B S, Caridad J M, Wang L, Hone J, Bøggild P and Booth T J 2016 *Nat Commun* **7** 11894
[39] Iwasaki T, Endo K, Watanabe E, Tsuya D, Morita Y, Nakaharai S, Noguchi Y, Wakayama Y, Watanabe K, Taniguchi T and Moriyama S 2020 *ACS Appl. Mater. Interfaces* **12** 8533
[40] Wu Z, Xu S, Lu H, Khamoshi A, Liu G B, Han T, Wu Y, Lin J, Long G, He Y, Cai Y, Yao Y, Zhang F and Wang N 2016 *Nat Commun* **7** 12955
[41] Movva H C P, Fallahazad B, Kim K, Larentis S, Taniguchi T, Watanabe K, Banerjee S K and Tutuc E 2017 *Phys Rev Lett* **118** 247701
[42] Larentis S, Movva H C P, Fallahazad B, Kim K, Behroozi A, Taniguchi T, Watanabe K, Banerjee S K and Tutuc E 2018 *Phys. Rev. B* **97** 201407
[43] Li L, Ye G J, Tran V, Fei R, Chen G, Wang H, Wang J, Watanabe K, Taniguchi

T and Yang L 2015 *Nat Nanotechnol* **10** 608

[44] Chen X, Wu Y, Wu Z, Han Y, Xu S, Wang L, Ye W, Han T, He Y, Cai Y and Wang N 2015 *Nat Commun* **6** 7315

[45] Li L, Yang F, Ye G J, Zhang Z, Zhu Z, Lou W, Zhou X, Li L, Watanabe K, Taniguchi T, Chang K, Wang Y, Chen X H and Zhang Y 2016 *Nat Nanotechnol* **11** 593

[46] Yuan K, Yin R, Li X, Han Y, Wu M, Chen S, Liu S, Xu X, Watanabe K, Taniguchi T, Muller D A, Shi J, Gao P, Wu X, Ye Y and Dai L 2019 *Advanced Functional Materials* **29** 1904032

[47] Gul O, van Woerkom D J, Weperen I, Car D, Plissard S R, Bakkers E P and Kouwenhoven L P 2015 *Nanotechnology* **26** 215202

[48] Guo Y, Wei X, Shu J, Liu B, Yin J, Guan C, Han Y, Gao S and Chen Q 2015 *Appl. Phys. Lett.* **106** 103109

[49] Hollinger G, Bergignat E, Joseph J and Robach Y 1983 *J. Vac. Sci. Technol. B* **1** 778

[50] Sun J, Lind E, Maximov I and Xu H Q 2011 *IEEE Electron Device Letters* **32** 131

[51] Dingle R B 1952 *Proc. R. Soc. Lond. A* **211** 517

[52] Coleridge P T 1991 *Phys Rev B Condens Matter* **44** 3793

[53] Das Sarma S and Stern F 1985 *Phys Rev B Condens Matter* **32** 8442

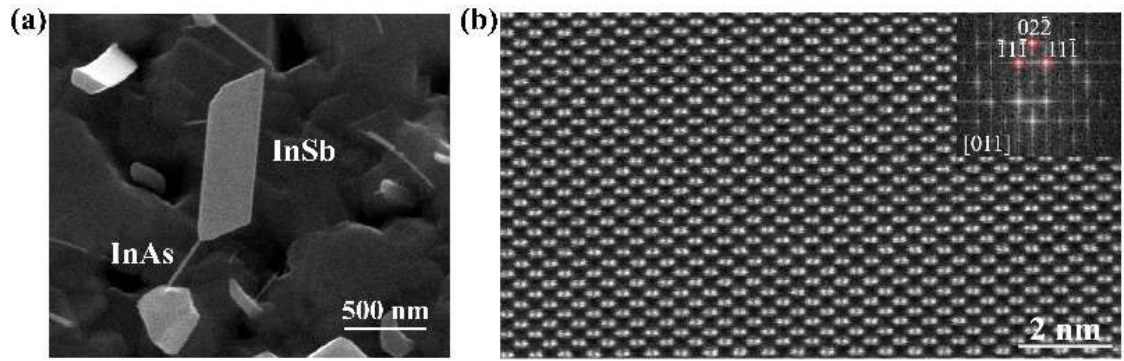

**Fig.1.** Crystal structure of an InSb nanosheet. (a) SEM image of an as-grown InSb nanosheet on the same growth Si(111) substrate as the InSb nanosheet used in this work. Here, the InAs nanowire, on which the InSb nanosheet was grown, is also shown. (b) HRTEM image of an InSb nanosheet from the same growth substrate. The inset is its FFT pattern. The HRTEM image and the FFT measurements show that the InSb nanosheet is a pure zincblende, single crystal with an atomic flat (011) top surface and is free from stacking faults and twinning defects.

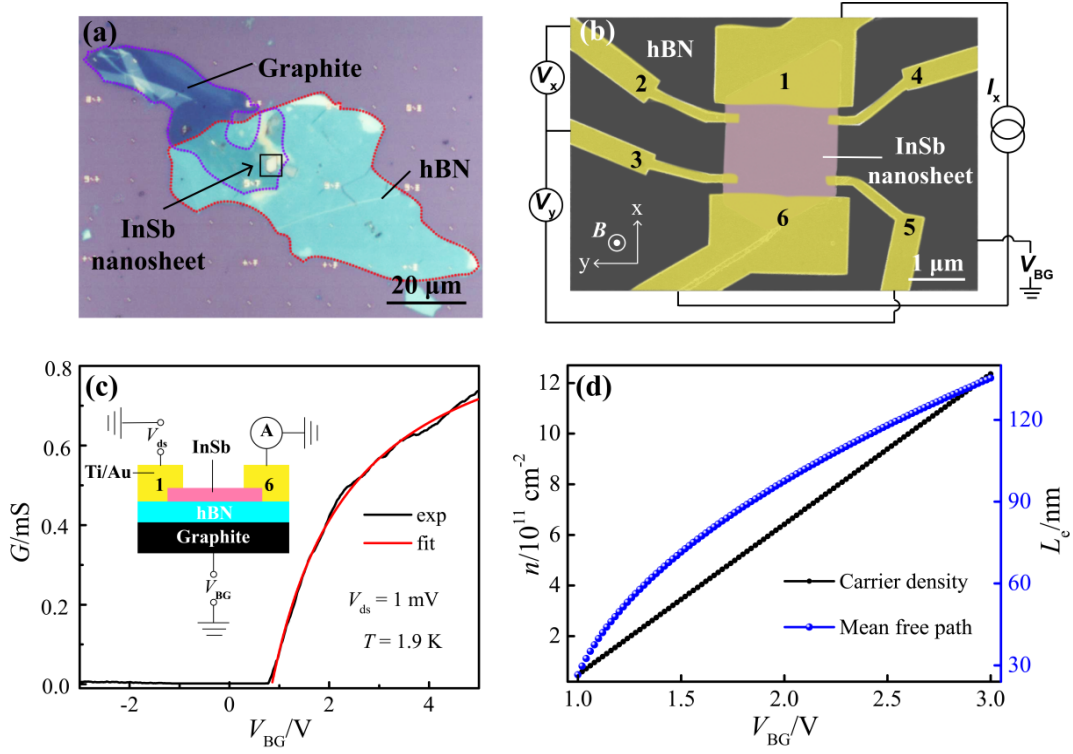

**Fig.2.** Device structure and gate-transfer characteristics. (a) Optical image of an InSb nanosheet/hBN/graphite heterostructure used for device fabrication in this work. (b) False-colored SEM image of an InSb nanosheet Hall-bar device studied by transport measurements in this work and schematics for magnetotransport measurement circuit setup. (c) Two-terminal DC conductance $G$ as a function of gate voltage $V_{BG}$ (gate-transfer characteristics) measured for the device shown in (b) at $V_{ds}$ = 1 mV and $T$ = 1.9 K (the black solid line). The red solid line presents the results of the fit of the on-state gate-transfer characteristic measurements to Eq. (1). The inset is a schematic cross-sectional view of the device and the circuit setup for the two-terminal gate-transfer characteristic measurements. (d) Carrier density $n$ and mean free path $L_e$ as a function of gate voltage $V_{BG}$ at the linear region of the gate-transfer characteristics shown in (c).

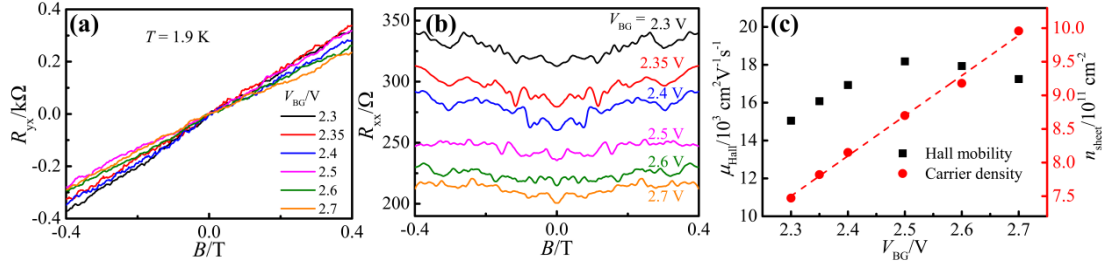

**Fig.3.** Low-field magnetotransport measurements and analyses. (a) Hall resistance $R_{yx}$ measured at different gate voltages and at $T = 1.9$ K as a function of magnetic field $B$ applied perpendicular to the InSb nanosheet plane. (b) Longitudinal resistance $R_{xx}$ measured at the same gate voltages as in (a) and at $T = 1.9$ K as a function of the magnetic field $B$. (c) Hall mobilities $\mu_{Hall}$ (black squares) and sheet carrier densities $n_{sheet}$ (red dots) vs. gate voltage $V_{BG}$ extracted from the low-field measurements shown in (a) and (b). The red dashed line shows the result of a linear fit to the $n_{sheet}$ vs. $V_{BG}$ data, from which a unit-area gate-to-channel capacitance of ~956 μF m$^{-2}$ can be extracted.

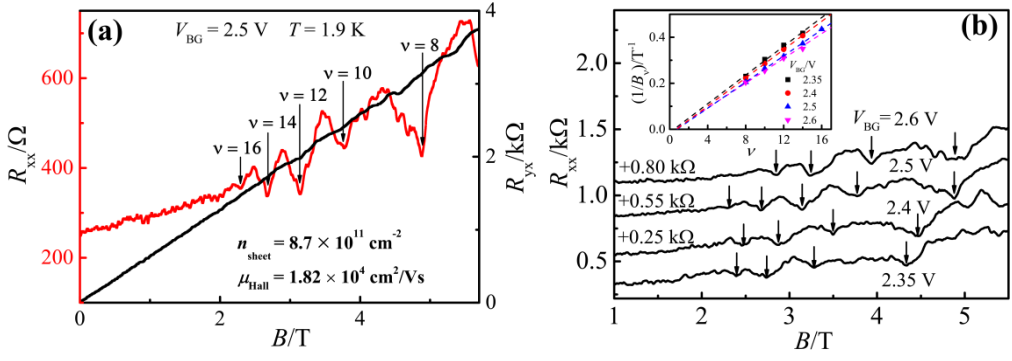

**Fig.4.** Gate dependent measurements of SdH oscillations in the InSb nanosheet. (a) Longitudinal resistance $R_{xx}$ (red solid line) and Hall resistance $R_{yx}$ (black solid line) vs $B$ measured at $T = 1.9$ K and $V_{BG}=2.5$ V ($n_{sheet} = 8.7 \times 10^{11}$ cm$^{-2}$). Here, SdH oscillations are observed in the measurements of $R_{xx}$. The black arrows mark the magnetic field positions at the bottoms of a few oscillation valleys with assigned filling factors $\nu$. (b) Longitudinal resistance $R_{xx}$ measured for the device at four different gate voltages as a function of magnetic field $B$. Here, the measured curves are offset vertically for clarity. The black arrows again mark the magnetic field positions at the bottoms of a few SdH oscillation valleys, at which the corresponding even filling factors $\nu$ are identified and are assigned. The inset shows the reciprocal values of the magnetic fields at the oscillation valleys $1/B_\nu$ vs. the assigned filling factors $\nu$ (symbols) at the four different gate voltages. The dashed lines in the inset are the straight line fits of the data points.

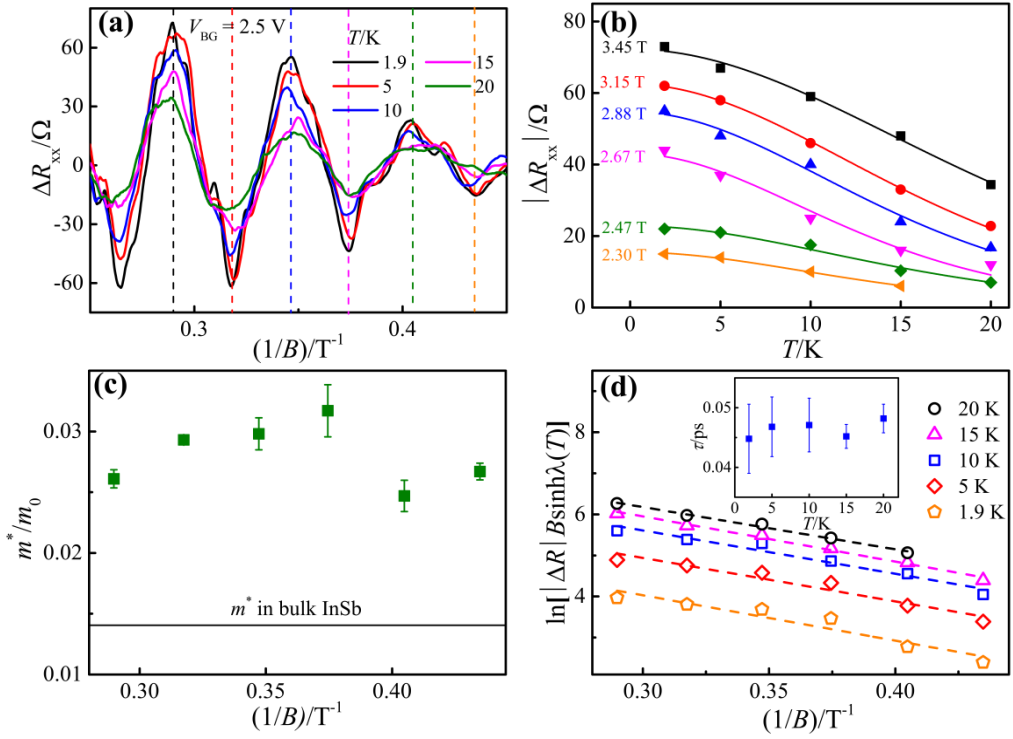

**Fig.5.** Cyclotron effective mass $m^*$ and quantum lifetime $\tau$ of the electrons in InSb nanosheet. (a) Longitudinal resistance $\Delta R_{xx}$, after subtracting a smooth background, as a function of $1/B$ at different temperatures but at fixed $V_{BG}$ = 2.5 V. (b) Amplitudes $|\Delta R_{xx}|$ of the SdH oscillations taken at the peaks and valleys marked by the vertical dashed lines in (a) as a function of temperature. The symbols are experimental data and the magnetic field values at which the oscillation amplitudes are taken are specified. The lines are the fits of the experimental data to the Lifshitz-Kosevich formula. (c) The electron effective mass extracted from the fits in (b). These extracted electron effective mass values give an average value of $m^*$=0.028 $m_0$ ($m_0$ is a free electron mass), which is larger than the effective mass value in bulk InSb (black solid line) and is expected due to the quantum confinement. Error bars represent standard deviations of the fits. (d) Experimental data of $\ln[|\Delta R|B\sinh\lambda(T)]$ (hollow symbols) plotted against $1/B$ at different temperatures (Dingle plots). The dashed lines are the linear fits of the experimental data. The inset shows the extracted quantum lifetimes from the fits at different temperatures, which give an averaged quantum lifetime of $\tau$ = 0.046 ps at $V_{BG}$ = 2.5 V (or $n_{sheet}$ = 8.7 × 10$^{11}$ cm$^{-2}$) in the InSb nanosheet.